\documentclass[%
 reprint,
superscriptaddress,
floatfix,
amsmath,amssymb,
 aps,
prb,longbibliography
]{revtex4-1}
\usepackage[utf8]{inputenc}
\usepackage{graphicx}
\usepackage{dcolumn}
\usepackage{bm}
\usepackage{color}
\usepackage{ulem}
\usepackage{todonotes}

\newcommand{\LCOO}{La$_2$CuO$_{4+y}$}
\newcommand{\LSCOO}{La$_{2-x}$Sr$_x$CuO$_{4+y}$}
\newcommand{\LSCO}{La$_{2-x}$Sr$_x$CuO$_{4}$}
\newcommand{\xiSize}{14}

\begin{document}
\title{Field-induced electronic phase separation in the \\ high-temperature superconductor La$_{1.94}$Sr$_{0.06}$CuO$_{4+y}$}


\author{S. Holm-Dahlin}
\email[]{sonja@nbi.ku.dk}
\affiliation{Nanoscience Center, Niels Bohr Institute, University of Copenhagen, 2100 Copenhagen {\O}, Denmark}

\author{J. Larsen}
\affiliation{Department of Physics, Technical University of Denmark, 2800 Kgs.\ Lyngby, Denmark}
\affiliation{Danish Fundamental Metrology, Kogle All\'e 5, 2970 H{\o}rsholm, Denmark}

\author{H. Jacobsen}
\affiliation{Nanoscience Center, Niels Bohr Institute, University of Copenhagen, 2100 Copenhagen {\O}, Denmark}
\affiliation{Department of Physics, Oxford University, Oxford, OX1 3PU, United Kingdom}
\affiliation{Laboratory of Neutron Scattering, Paul Scherrer Institute, 5232 Villigen PSI, Switzerland}

\author{A. T. R\o mer}
\author{A.-E. \c{T}u\c{t}ueanu}
\affiliation{Nanoscience Center, Niels Bohr Institute, University of Copenhagen, 2100 Copenhagen {\O}, Denmark}
\affiliation{Institute Max von Laue Paul Langevin, 38042 Grenoble, France}
\affiliation{Danish Fundamental Metrology, Kogle All\'e 5, 2970 H{\o}rsholm, Denmark}

\author{M. Ahmad}
\affiliation{Nanoscience Center, Niels Bohr Institute, University of Copenhagen, 2100 Copenhagen {\O}, Denmark}

\author{J.-C. Grivel}
\affiliation{Department of Energy Conversion, Technical University of Denmark, 2800 Kgs.\ Lyngby, Denmark}

\author{R. Scheuermann}
\affiliation{Laboratory for Muon Spin Spectroscopy, Paul Scherrer Institute, 5232 Villigen PSI, Switzerland}

\author{M. v. Zimmermann}
\affiliation{Deutsches Elektronen-Synchrotron DESY, Notkestr. 85, 22603 Hamburg, Germany}

\author{M. Boehm}
\author{P. Steffens}
\affiliation{Institute Max von Laue Paul Langevin, 38042 Grenoble, France}

\author{Ch. Niedermayer}
\affiliation{Laboratory of Neutron Scattering, Paul Scherrer Institute, 5232 Villigen PSI, Switzerland}

\author{K. S. Pedersen}
\affiliation{Department of Chemistry, Technical University of Denmark, 2800 Kgs.\ Lyngby, Denmark}

\author{N. B. Christensen}
\affiliation{Department of Physics, Technical University of Denmark, 2800 Kgs.\ Lyngby, Denmark}

\author{B. O. Wells}
\affiliation{Department of Physics and Institute of Materials Science, Univ. Connecticut, Connecticut 06269, USA}

\author{K. Lefmann}
\email[]{lefmann@nbi.ku.dk}
\author{L. Udby}
\affiliation{Nanoscience Center, Niels Bohr Institute, University of Copenhagen, 2100 Copenhagen {\O}, Denmark}

\date{\today}

\begin{abstract}
We present a combined neutron diffraction and high field muon spin rotation ($\mu$SR) study of the magnetically ordered and superconducting phases of the high-temperature superconductor La$_{1.94}$Sr$_{0.06}$CuO$_{4+y}$ ($T_{\rm c} = 37.5(2)$~K) in a magnetic field applied perpendicular to the CuO$_2$ planes. 
We observe a linear field-dependence of the intensity of the neutron diffraction peak that reflects the modulated antiferromagnetic stripe order. The magnetic volume fraction extracted from $\mu$SR data likewise increases linearly with applied magnetic field.
The combination of these two observations allows us to unambiguously conclude that stripe-ordered regions grow in an applied field, whereas the stripe-ordered magnetic moment itself is field-independent. This contrasts with earlier suggestions that the field-induced neutron diffraction intensity in La-based cuprates is due to an increase in the ordered moment. We discuss a microscopic picture that is capable of reconciling these conflicting viewpoints.
\end{abstract}

\maketitle

\color{black}

\section{Introduction}

Magnetic fluctuations are the most likely pairing mechanism behind high-temperature superconductivity (SC) in the cuprates \cite{Scalapino2012}, and therefore deserves to be studied in detail. The static antiferromagnetic (AFM) order of the parent Mott insulator compounds prevents SC from occurring. Upon doping, the AFM order is suppressed and SC appears, however, charge density wave order, observed by x-ray diffraction studies, also emerges upon doping and is likewise known to compete across several cuprate families with SC\cite{Ghiringhelli821,Chang2012,Thampy2014}.

The earliest direct evidence of modulated charge and magnetic ordering in cuprate superconductors was the discovery of \textit{stripe} order in hole-doped La$_{2-x-y}$Nd$_{y}$Sr$_{x}$CuO$_4$ \cite{Tranquada1995} found with neutron diffraction studies and later in La$_{2-x}$Ba$_x$CuO$_4$ \cite{Fujita2004}, in La$_{2-x-y}$Eu$_{y}$Sr$_{x}$CuO$_4$ \cite{Hucker2007,Fink2009}, and in La$_{2-x}$Sr$_x$CuO$_4$\cite{Suzuki1998}, all of them chemically a-site doped La-214 type cuprates. 

Another route for introducing holes to the CuO$_2$ planes is adding excess-oxygen to the material, {\it e.g.} in YBa$_2$Cu$_3$O$_{6+y}$\cite{Wang1987} and La$_{2}$CuO$_{4+y}$ \cite{Wells1997}. Unlike doping by chemical substitution, the oxygen dopants remain mobile in \LCOO{} down to temperatures around 200~K \cite{Xiong1996,Lee2004,Fratini2010}. Above this temperature, the holes can distribute in a way that at low temperatures favors long-range ordered stripe regions with a period of 8 unit cells \cite{Lee1999} and support connected SC regions to give the highest critical temperature $T_{\rm c}$ for the La-214 family \cite{Fratini2010}. 

In many oxygen stoichiometric La-214 cuprates, stripe order is most prominent at dopings just around $x=0.125$, accompanied by a suppression of $T_{\rm c}$ \cite{Moodenbaugh1988, Fujita2004, Tranquada2008, Hucker2011, Suzuki1998, Chang2008}. Here, the intensity of the neutron diffraction (ND) peak from the stripes is enhanced by an applied magnetic field \cite{Lake2002,Chang2008, Roemer2013}. At higher doping levels, the stripe order is absent, but can be induced by a magnetic field \cite{Khaykovich2005,Chang2008, Chang2009, Frachet2020}.

The field-driven enhancement of the magnetic ND peak has often been interpreted as a field-induced increase of the ordered moment in La-214 cuprates \cite{Lake2002,Khaykovich2002,Chang2008}. However, studies using neutron diffraction and muon spin rotation ($\mu$SR) have revealed that oxygen doped La-214 cuprates exhibit \textit{electronic phase separation} between optimally doped SC without static magnetism and a magnetic stripe ordered phase without SC \cite{Savici2002,Kofu2009, Kivelson2001}. This finding stands in contrast to the suppression of SC found in oxygen stoichiometric La-214 mentioned above. 

In co-doped La$_{2-x}$Sr$_x$CuO$_{4+y}$, holes are introduced both by chemical substitution and by intercalating extra oxygen. So far, stripe order is found in all oxygen co-doped samples when they are optimally oxygen-doped, achieving $T_{\rm c} \approx 40 $~K  \cite{Lee1999,Mohottala2006,Udby2013,Jacobsen2018}. In contrast, oxygen-stoichiometric \LSCO{} shows stripe order only in the underdoped region, $x<0.135$, with $T_{\rm c}$ of 30~K or less \cite{Wakimoto2001,Chang2008,Kofu2009,Khaykovich2005,Chang2009}. 

Detailed transport studies of La$_{1.875}$Ba$_{0.125}$CuO$_4$ uncovered evidence for 2D superconducting correlations developing at the spin-ordering temperature, $T_N$, in zero field, but becoming rapidly suppressed upon application of a magnetic field perpendicular to the CuO$_2$ planes \cite{Li2007}. Evidence for electronically decoupled planes associated with stripe magnetic order has likewise been found in optical studies of underdoped, stripe-ordered La$_{2-x}$Sr$_{x}$CuO$_4$ and La$_{2-x-y}$Nd$_{y}$Sr$_{x}$CuO$_4$ superconductors \cite{Tajima2001,Schafgans2010_1,Schafgans2010_2}. 
In the case of La$_{2-x}$Sr$_{x}$CuO$_4$, application of a c-axis magnetic field suppresses the Josephson plasma resonance observed by infrared spectroscopy, \cite{ Schafgans2010_1} interpreted as a weakening of the interlayer Josephson coupling. 
The observations of 2D superconductivity in the cuprates inspired the development of theories for pair-density wave (PDW) order \cite{Himeda2002,Berg2007,Agterberg2020} and more broadly the concept of intertwined orders in high-$T_c$ superconductors \cite{Fradkin2015}.  

It is evident that there is a rich interplay between the 2D SC, the stripe order, both structural and magnetic, and 3D SC in these cuprate compounds; And that there is a need to investigate the different phases in the cuprates in order to understand the competition, interplay, and phases separation of the different states of matter.

In this work, we study the magnetic stripes in co-doped  La$_{1.94}$Sr$_{0.06}$CuO$_{4+y}$. We use $\mu$SR to show that there is no static magnetism in our sample, in contrast to \LSCOO{} compounds with other Sr dopings \cite{Mohottala2006,Udby2013}. We further show that an applied magnetic field induces the magnetic order, increasing the magnetic volume fraction linearly from 0\% at 0~T to $37(3)$\% at 8~T. At the same time, the intensity of the ND peak from the stripe order also increases linearly with an applied magnetic field. The combination of these observations implies that the field-induced ND peak is caused solely by the increase in the magnetic volume fraction, while the size of ordered magnetic moment is field-independent. This is in contrast to some conclusions drawn in earlier studies, where the ordered magnetic moment was found to increase with increasing field \cite{Lake2002,Khaykovich2002,Chang2008}.

\section{Methods}

A 7.9~g single crystal was grown by the travelling solvent floating zone method and was oxygenated through a wet-chemical technique \cite{Mohottala2006} obtaining a single onset transition temperature of $T_{\rm c}=37.5(2)$~K. Magnetization measurements were performed both before and after oxidation, see Appendix A. To ensure consistency between our experiments, the sample was always field cooled; The sample was heated to at least 50~K (well above $T_c$ and the Néel temperature $T_N$) before any change of applied magnetic field. We adopted a slow cooling procedure as described in Ref.~\onlinecite{Lee2004} to allow the oxygen to find an optimal configuration before freezing \cite{Lorenz2002,Fratini2010}, see Appendix A for details.

ND was performed at the cold triple axis spectrometers RITA-II at Paul Scherrer Institute (PSI), Switzerland \cite{bahlrita1,bahlrita2} and ThALES at Institute Laue-Langevin (ILL), France \cite{boehm07,ILL_data,Lefmann2015}. In both ND experiments, the sample was mounted in a 14.9~T cryo-magnet with the $c$-axis along the field. The incoherent elastic energy resolution was $\sim 0.2$~meV (full width at half maximum). The crystal was mounted to allow access to wave vectors ${\bf Q}=(h,k,0)$, expressed in units of $(2\pi/a,2\pi/b,0)$ (using orthorhombic notation), see Appendix D for details on the instrument setup. The sample configuration allowed us to measure the peaks from the stripes, which appears as a quartet of weak peaks around the antiferromagnetic point ${\bf Q_{\rm AFM}}=(1,0,0)$ \cite{Yamada1998}. 

To measure the inter-planar magnetic correlations, the sample was mounted with the $c$-axis in the scattering plane in a horizontal field magnet. The experiment was carried out at the RITA-II spectrometer with the same configuration as described above (and in Appendix D).

The $\mu$SR measurements were carried out at the General Purpose Spectrometer \cite{GPS} and the high transverse field spectrometer HAL9500 \cite{HAL9500}, both located at PSI. We used a smaller piece of the same sample and applied the magnetic field along the $c$ axis. See Appendix C for details on the $\mu$SR measurements and data analysis.

We performed hard X-ray diffraction experiments at the BW5 beamline at HASYLAB, Deutsches Elektronen-Synchrotron, Hamburg, Germany. Here we looked for a signature of charge-density waves. The sample was cut into a thin slice ($d = 1.4$~mm) in order to reduce the absorption, meaning that 10 \% of the incoming beam was transmitted through the sample.

The magnetization measurements were carried with the magnetic field applied along the $c$-axis in a vibrating sample magnetometer (VSM) in a Physical Properties Measurement System (PPMS) located at the Technical University of Denmark.

\section{Results}

\begin{figure}[h]
\centerline{
\includegraphics[width=0.5\textwidth]{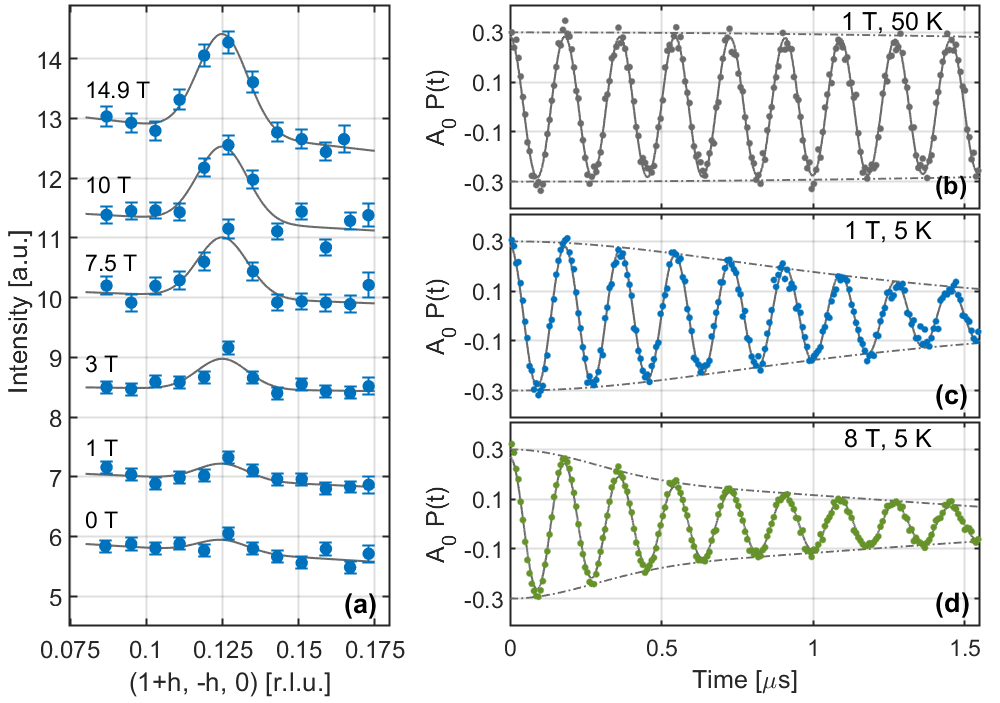} 
} \caption{(a): ND data measured at RITA-II at 1.6~K, counting between 7 and 18 min. per point. Each data set is offset by a multiple of 1.5~a.u.. Solid lines are Gaussian fits, as described in the text. (b-d): Examples of $\mu$SR asymmetry data in applied magnetic fields measured at HAL9500. Solid lines are fits as described in the text, and dashed envelopes highlight the difference in the fitted relaxation rates.
} \label{fig:rawdata}
\end{figure}

A selection of raw ND data is presented in Fig.~\ref{fig:rawdata} (a), showing a clear field-induced peak at 14.9~T. The intensity of the peak decreases with decreasing field, but a tiny signal may still be present at 2~K in zero field. However, the error on the fitted amplitude in zero field is nearly as large as the value itself. We fitted the data using Gaussian peaks on a sloping background. Due to the weak signal, not all parameters were fitted simultaneously. The peak center and width show no visible dependence on the magnetic field, and we have thus fixed these to the values obtained at 14.9~T. 

The stripe peaks are present at all four scattering vectors  ${\bf Q_{\rm stripe}}=(1\pm\delta_h,\pm\delta_k,0)$ with $\delta_h\approx \delta_k = 0.124(2)$, and corresponds to a period-8 antiferromagnetic modulation along the Cu-O bonds. The peak located at $(1+\delta_h,-\delta_k,0)$ was used for the temperature and magnetic field dependence studies. The full width at half maximum of the peak is $2w = 0.019(2)$ r.l.u., which is similar to measurements on other La-214 cuprates at similar cold neutron instruments, see {\it e.g.} Ref.~\onlinecite{Chang2008}. 

The intrinsic width of the peak ($\gamma$) is related to the correlation length of the stripe regions, $\xi_{\rm AFM} = 1/\gamma$. To determine $\xi_{\rm AFM}$, we need to account for the instrument resolution, $\sigma$. Since the resolution function depends on the scattering angle, the resolution near the forbidden (100) peak cannot be determined experimentally from measurements of other peaks, such as the (200) nuclear Bragg peak \cite{Shiranebook}. We have therefore estimated the instrument resolution for the particular scan in two ways: (a) Simulating  an ideal sample using McStas\cite{Willendrup2020}, which is known to reproduce line widths to a precision of a few percent \cite{Udby2011, bahlrita2}, yielding $\sigma=0.0147$ r.l.u. and (b) measuring the second order scattering on the nominal (100) peak, yielding  $\sigma=0.0170$ r.l.u. From these two estimates, the intrinsic width of the peak is (a) $\gamma=0.012(3)$ r.l.u or (b) $\gamma=0.009(12)$ r.l.u (i.e., not significantly different from 0). A conservative lower bound on the correlation length of the stripe regions from estimate (a) is $\xi_{\rm AFM} = 1/\gamma > \xiSize{} $~nm, while estimate (b) is  consistent with long range order.

Fig.~\ref{fig:rawdata} (b-d) show a selection of $\mu$SR asymmetry spectra in a rotating reference frame (RRF) chosen such that the very high muon spin precession frequencies at high fields are scaled down for better visualization of the data. The RRF frequency is 130~MHz at 1~T and 1080~MHz at 8~T. In panel (b), data taken at 50~K shows an undamped oscillation of the muon spin asymmetry that is well described by a model including a single Gaussian component with a frequency corresponding to the external magnetic field of 1~T. This behavior is expected for a paramagnetic sample. At 5~K and 1~T in panel (c), the spectrum shows a clear decay of the muon spin asymmetry, which is dominated by one oscillating component with Gaussian relaxation. In panel (d) at 8~T and 5~K, the time evolution of the muon spin asymmetry is best described using two distinct components corresponding to the magnetic and non-magnetic regions in the sample, respectively. Below, we argue that the non-magnetic regions of the sample are superconducting. The pronounced rapid decay of the muon asymmetry at early times is modelled by a Gaussian oscillation with an enhanced relaxation rate, consistent with muons stopping in magnetically ordered regions. The magnetic and SC components have rotation frequencies, $\omega_\text{m}$ and $\omega_\text{SC}$, and Gaussian decay parameters $\sigma_\text{m}$ and $\sigma_\text{SC}$, respectively. For details on the $\mu$SR data treatment, see Appendix C.

\begin{figure}[t]
\includegraphics[width=0.45\textwidth]{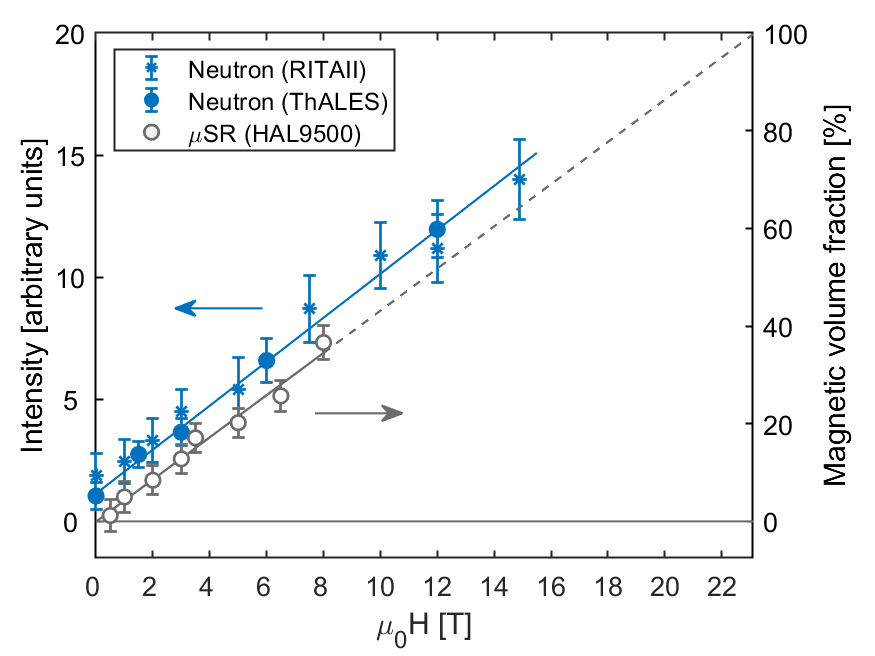}
\caption{The low temperature magnetic field dependence of the neutron diffraction intensity from the stripe phase (blue) and the magnetic volume fraction  measured with $\mu$SR (gray). 
} \label{fig:linear}
\end{figure}

Fig.~\ref{fig:linear} shows the main result of this study; the linear dependence of the ND intensity and magnetic volume fraction ($V_{\rm m}$) found with $\mu$SR measured at low temperatures (RITA-II at 1.8~K, ThALES at 1.5~K, and HAL9500 at 2.5~K). 
The linear dependence has been found in two different neutron experiments, and their axes have been scaled independently to have the linear fits coincide on the arbitrary scale. 
The $\mu$SR measurements are on an absolute scale and directly show that the magnetically ordered fraction of the sample scales linearly with applied field. 

\begin{figure}[h]
\includegraphics[width=0.47\textwidth,trim=0 0mm 0 0, clip]{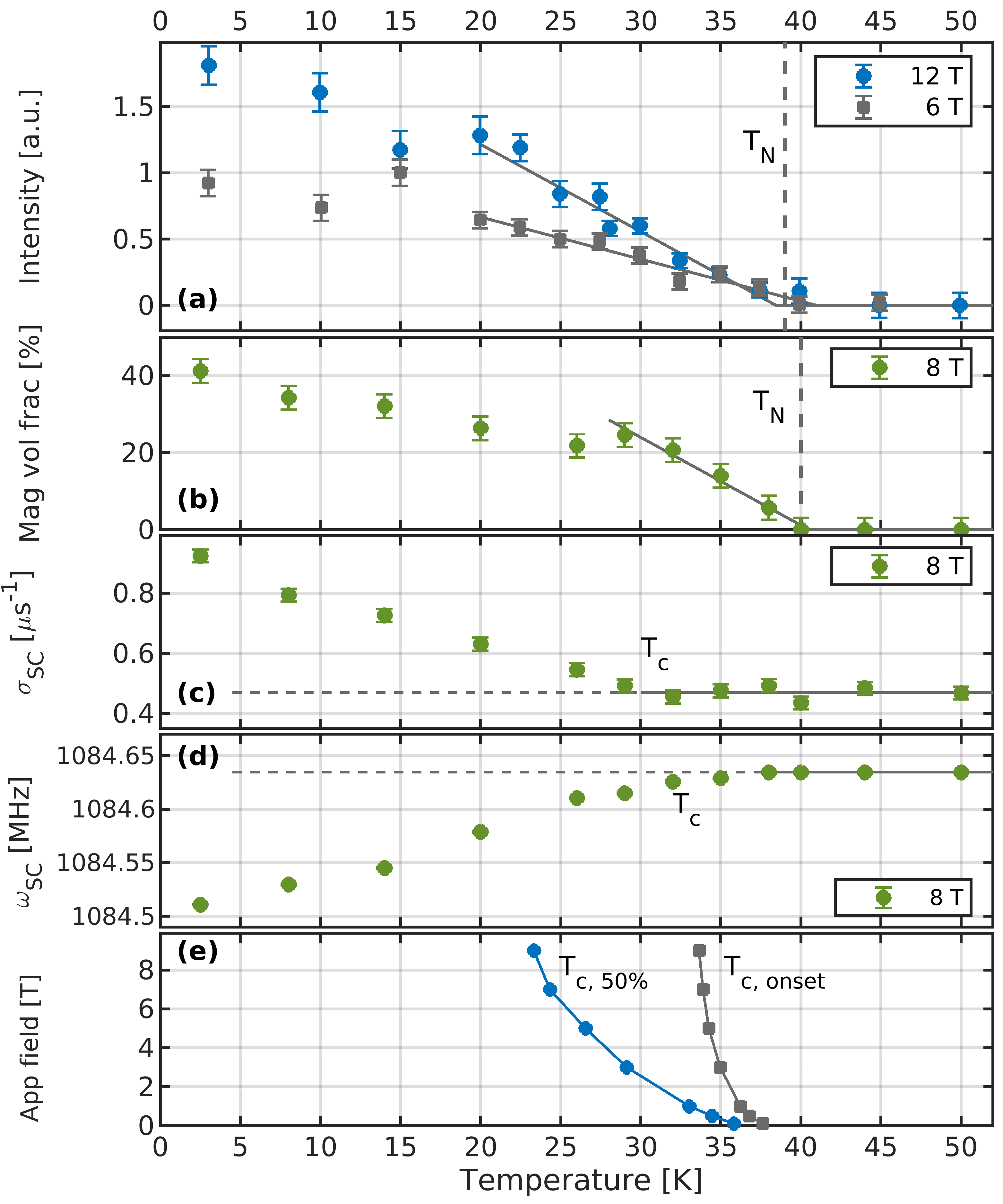} 
\caption{
(a): The intensity of the ND peak as a function of temperature for applied fields of 6 T and 12 T. The solid lines are linear fits to the 6~T and 12~T data in the range 20-40~K. (b-d): $\mu$SR fitting parameters determined in 8~T. The solid line in (b) is a linear fit to $V_{\rm m}$ in the range 30-40~K. The constant lines in (c) and (d) are fits to $\sigma_\text{SC}$ and $\omega_\text{SC}$ above 38~K.
(e): $T_{\rm c}$ onset and 50\% magnetization point as a function of field, found from field-cooled magnetization measurements.
}
\label{fig:temperature}
\end{figure}

In Fig.~\ref{fig:temperature} the temperature dependence of parameters found with three different techniques are displayed together. Fig.~\ref{fig:temperature} (a) shows the integrated intensity of the ND peak measured at ThALES. The magnetic ordering temperature is determined to $T_{\rm N}= 39(1)$~K, for both 6~T and 12~T. 

Fig.~\ref{fig:temperature} (b) displays $V_{\rm m}$ as a function of temperature, extracted from the $\mu$SR data. The fast relaxation associated with magnetic order vanishes at 40(1)~K, consistent with $T_{\rm N}= 39(1)$~K found with neutrons. Panel (c) shows that the relaxation rate $\sigma_\text{SC}$ of the non-magnetic regions takes a constant value of 0.47 $\mu$s$^{-1}$ at higher temperatures. On cooling below 32~K, $\sigma_\text{SC}$ increases, to reach a value of 0.95(5) $\mu$s$^{-1}$ at 2~K.
In panel (d) the rotation frequency of the muons in the non-magnetic regions is seen to be constant at high temperature, with a value that corresponds to the external magnetic field. The small negative shift of $\omega_\text{SC}$ below 38~K together with the increase of  $\sigma_\text{SC}$ appearing below $T_{\rm c}$ is typical for SC and can be understood in terms of a broadening of the field distribution due to the formation of a flux line lattice within the SC state that forms below $T_{\rm c}$ \cite{Blundell1999}. We thus ascribe this non-magnetic component at low temperatures to muons stopping in SC regions of the sample, as done in Refs.~\onlinecite{Mohottala2006, Udby2013}. There is no evidence of a third component that is simultaneously non-SC and non-magnetic.

\begin{figure}[h]
\includegraphics[width=0.45\textwidth]{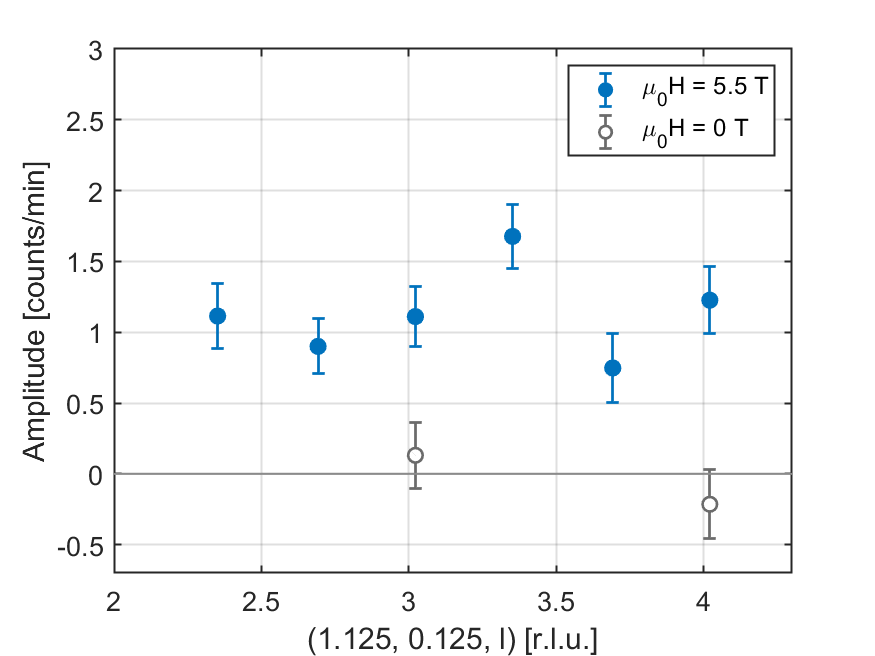} 
\caption{Neutron diffraction intensities shows the two-dimensional nature of the magnetic peak along the $l$-direction, measured at 2.3~K.} \label{fig:ldep}
\end{figure}

The temperature dependence of the sample magnetization is shown in Fig.~\ref{fig:temperature} (e). We measured for 7 different field values in the range $0.1 - 9$~T. The transition from the normal to the SC state broadens drastically with applied field. The raw data and fits are shown in Fig.~\ref{SC_rawMdata} in Appendix B. We determine the onset of the SC transition $T_{\rm c, onset}$ as well as the 50\% magnetization point $T_{\rm c, 50\%}$ as a function of field with linear fits to the magnetization data, see Fig.~\ref{SC_rawMdata} in Appendix B. 
From the field dependence of $T_{\rm c}$ in low fields, it is possible to give an estimate of the critical field, $H_{c2}(0)$, and thereby the superconducting coherence length, $\xi$, using the Werthamer-Helfand-Hohenberg (WHH) model. This crude estimate gives $H_{c2}(0)\approx 17-54$~T and $\xi\sim2.5-4.5$~nm in agreement with other dopings of \LSCO{} \cite{Wang2008a}. See Appendix B for details on the WHH model.

\begin{figure}[b]
\includegraphics[width=0.42\textwidth]{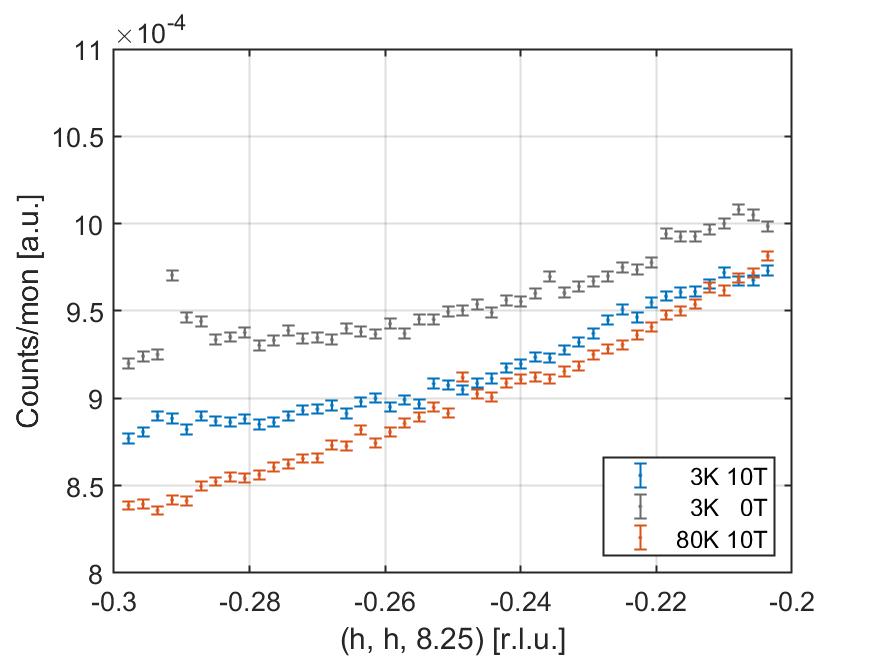} 
\caption{Examples of the raw data from the X-ray diffraction experiments. No field-induced diffraction peak indicating charge density waves can be found in the data measured at 3~K and 10~T (gray points).} \label{fig:charge}
\end{figure}

In Fig.~\ref{fig:ldep} the $l$-dependence of the amplitude of the magnetic peak is shown. Sample rotation scans were performed above $T_{\rm N}$ and subtracted as background from the low temperature data at 2.3~K. This procedure was performed in zero field and in an applied field of 5.5~T. The data was binned with a step size of 0.3~r.l.u. along $l$. 
Due to the mismatch between the poor out of plane resolution of a triple-axis-instrument and the rod-like signal along $l$ from the magnetic scattering, it is extremely difficult to get adequate statistics for this experiment. Moreover, the maximum field strength of the horizontal field magnet is 5.5~T, and only about 20\% of the sample is in the magnetic phase at this field. With a counting time of 48 minutes per point, it was only possible to extract the amplitudes of Gaussian fits with fixed values of the width and center (the background is zero due to the subtraction). Under these conditions, the data shows a field-induced magnetic signal but no intensity modulation along $l$. 

We searched for a field-induced X-ray diffraction signal at $(\pm0.25, \pm0.25, l)$ for several values of $l=$ 8, 8.25, 8.5, 9.5, 11.5, and 12.5, in an applied field of 10~T. Even with an extreme counting time of up to 7.5 minute per point, no signal was found. Fig.~\ref{fig:charge} shows an example of our raw data.

\section{Discussion}

The onset of the magnetic signal is found, within error, at the same temperature, independent of applied field (6~T, 8~T, and 12~T) measured with both ND and $\mu$SR, see Fig.~\ref{fig:temperature} (a-b). At 8~T, the SC transition ($T_{\rm c}$ onset) found with magnetization measurements coincides with the onset temperature found by $\mu$SR at 8~T as seen in Fig.~\ref{fig:temperature} (e) compared to (c-d). 
At low fields, the SC and magnetic phases have the same transition temperature, but at higher fields the two transition temperatures differ from each other, as $T_{\rm N}$ remains unchanged while $T_{\rm c}$ is suppressed with field. This result is consistent with findings in \LSCO{} with $x=0.10$ having $T_{\rm N} \sim T_{\rm c}(H=0)=30$~K \cite{Lake2002}.
While the similarity of $T_{\rm N}$ and $T_{\rm c}$ at low fields might be seen as an indication of one correlated phase, we interpret the different onset temperatures at higher field values as a signature of two competing phases, magnetic and SC, with almost degenerate free energy in zero field. The application of a magnetic field favors the formation of regions with quasi-static stripe order and suppresses $T_{\rm c}$.

Extrapolating our $\mu$SR results to higher fields in Fig.~\ref{fig:linear}, $V_{\rm m}$ reaches 100\% at $B_{\rm sat} \sim 23$~T, suppressing the SC phase completely. This is consistent with our crude estimate of $H_{c2}(0)\approx 17-54$~T from the WHH model. This further indicates that the magnetic stripe phase and SC competes. 
It is, however, remarkable that we find no evidence of charge stripes of the type observed in other La-214 compounds, even though we used the exact same method as in Refs.~\onlinecite{Christensen2014,Croft2014}. This finding is consistent with a picture of magnetic fluctuations rather than static magnetic (and structural) order. In Ref.~\onlinecite{Zhang2018} it is discussed how the formation of static charge stripes happens, and that the relationship between charge nematicity and the magnetic stripe signal may not be simple.

In zero field a tiny ND signal seems to be present, while no static magnetism is observed with $\mu$SR. The small discrepancy between the results may be explained by the difference in timescale between the two techniques. Cold neutron measurements probe a timescale of $\sim 10^{-11}$~s and can not distinguish fluctuations on longer timescales from static magnetism. In contrast, high transverse field $\mu$SR measurements probe a timescale of $\sim 10^{-7}$~s. Magnetic moments fluctuating on a timescale between these two values would thus be seen as static by the neutrons, but as fluctuating by the muons. The difference between the two techniques may thus be reconciled if the stripe signal is not truly static, but rather slow fluctuations, as discussed e.g. in Refs. \onlinecite{Wakimoto1999,Roemer2013,Jacobsen2015}. Therefore, the peak at 0 T likely does not originate from truly static moments, but low-energy fluctuations picked up by the finite energy resolution of the instrument.

The ND peak at low temperature increases linearly with magnetic field, see Fig.~\ref{fig:linear}. The measured ND intensity scales as $I \propto V_{\rm m} M_\perp^2$, where $V_{\rm m}$ is the magnetic volume fraction and $M_\perp$ is the component of the ordered moment perpendicular to the scattering vector \cite{Shiranebook}. At the same time, our $\mu$SR experiments show a pure SC phase at zero field and a linear increase of $V_{\rm m}$ with applied field. Combining the results of the two techniques, constituting a local probe and a coherent probe, we conclude that the observed increase in the ND signal is caused purely by an increase in  $V_{\rm m}$, and not by an increase of $M_\perp$.

Previous ND investigations of \LSCO{} in samples that exhibit near 100\% magnetic volume fraction in zero field \cite{Chang2008,Roemer2015} found a field-enhanced intensity, which was ascribed to an increase in the ordered moment of pre-existing magnetic order. A similar interpretation was presented for \LCOO{} in Ref.~\onlinecite{Khaykovich2002}. In contrast, the linear increase in the magnetic scattering intensity in our sample can be fully accounted for by the linear increase in the magnetic phase fraction derived from the muon signal, leaving no evidence for an increase in the size of the local moments. 
This raises an apparent contradiction. While the phenomenology of the field-enhanced scattering in the two systems seem incompatible with a single mechanism, it seems unlikely that the effect of a magnetic field in two such similar La-214 systems is fundamentally different. Below we present an interpretation of the available data which allows for a single description favoring the phase fraction description. 

There are only few studies of the field dependent magnetic scattering that include both $\mu$SR and neutron diffraction data. This makes it difficult to construct a direct connection between phase fractions and the magnetic scattering intensity. An exception is the combined ND and $\mu$SR study of \LSCO{} samples of $x=0.105$, 0.12, and 0.145 in Ref.~\onlinecite{Chang2008}. Of particular note is the \LSCO{} sample with $x=0.12$ which shows significant field enhanced scattering but near 100\% magnetic phase fraction even in zero field. However, $\mu$SR does not measure local moments directly, but the local magnetic field, which is influenced by moments arrayed over a range of around 10-20 \AA. Ref.~\onlinecite{Chang2008} notes that the magnetic structure of the $x=0.12$ sample may be inhomogeneous at this length scale. 

Thus, a possible connected picture emerges. In the \LSCOO{} sample reported here, phase separation creates large magnetic and SC regions, and the application of a magnetic field increases the magnetic volume fraction in a straightforward manner. In the \LSCO{} sample with $x=0.12$, a frustrated tendency towards phase separation leads to inhomogeneity on a sub-nanometer scale. Application of a field again favors formation of the striped magnetic region, but in this case the induced stripe phase fills the spatial gaps between existing striped regions, creating a more homogeneous phase and larger local fields at the muon sites.  

A similar explanation of a field-induced shift in balance between the volume of two competing volume fractions could be relevant also for the superconductor YBa$_2$Cu$_3$O$_{6+y}$, where $T_N\simeq T_c$ and a linear field-dependence of the magnetic neutron diffraction peak is observed \cite{Haug2009}.

In general, there is experimental evidence that magnetic stripes co-exist with SC in 
the La-214 materials\cite{Tranquada1995,Christensen2014,Croft2014,Thampy2014}, but it remains unknown if the coexistence is microscopic or not. 
The superconducting coherence length in this sample has been estimated via the WHH model to be in the range $\xi=2.5 - 4.5$~nm, which is in agreement with other \LSCO{} compounds, e.g. Ref.~\onlinecite{Wang2008a}. We find the lower limit of the magnetic correlation length to be significantly larger, $\xi_{\rm AFM}>\xiSize$~nm. These findings suggest that the field-induced stripe order is correlated beyond the size of a vortex core, in agreement with e.g. Ref.~\onlinecite{Lake2002}. Therefore, the simple picture that magnetism is found only inside the vortex core is inadequate, and further investigations of the overlap between the magnetic and SC phases are needed. 

When discussing domain sizes, it is important to note that these systems are somewhat two-dimensional in nature. There is a striking parallel between the phenomenology of broken interlayer Josephson coupling detected optically, and that of the field-dependence of stripe magnetic order detected by neutron diffraction: When stripe order is well-developed and $T_c$ strongly suppressed, as in the case of La$_{2-x}$Ba$_x$CuO$_4$ and La$_{2-x-y}$Nd$_{y}$Sr$_{x}$CuO$_4$ near $x=1/8$, 2D superconducting correlations are detected in zero magnetic field \cite{Schafgans2010_2}, while the incommensurate stripe magnetic order is field-independent \cite{Chang2008,Wen2008}. On the other hand, stripe order is less well-developed under zero field conditions in underdoped \LSCO{}, and $T_c$ correspondingly less suppressed. In this family of materials, diffraction studies indicate field-enhanced Bragg peak intensities \cite{Lake2002,Chang2008}, and as stripe order strengthens optical studies indicate a gradual emergence of 2D superconductivity \cite{Schafgans2010_1,Schafgans2010_2}. It therefore appears that stripe order and 2D superconducting correlations go hand in hand.

The intensity of the ND signal from magnetic stripes shows no or little variation along the out-of-plane $l$-direction. This demonstrates that the field-induced magnetic peak in our sample is two-dimensional, different from the zero-field magnetic signal in \LSCO{} where short-ranged magnetic correlations along the $c$-axis were found for $x=0.10$ \cite{Lake2005} and $x=0.12$ \cite{Roemer2015}. Also in \LCOO{} pronounced modulations along $l$ have been found \cite{Lee1999}. It is, however, possible that any modulation of the neutron signal from a single (magnetic) domain could be washed out, and thus remain unobserved, due to a scattering signal from other magnetic domains (e.g. due to twinning in the crystal) with large intensity at different $l$-values \cite{Roemer2015}. Thus, observation of (short-range) magnetic c-axis correlations could depend on systematic or accidental detwinning of particular samples.

Some of us earlier studied other co-doped samples of La$_{2-x}$Sr$_{x}$CuO$_{4+y}$ with values of the Sr-doping $x$ being 0.04, 0.065, 0.09, and 0.115.  For the sample with $x=0.09$, we found no field-enhancement of the ND stripe signal \cite{Udby2009}. This could suggest that parts of the $x=0.09$ sample would display 2D superconductivity, while other parts would be stripe-free and 3D superconducting.

The magnetic volume fraction in zero field for co-doped samples with different Sr content vary between 20\% and 65\% \cite{Mohottala2006}. Our sample, however, has no magnetic volume fraction in zero magnetic field. 
One can speculate that our sample is positioned at the cross-over point between annealed order (O-doping) and quenched disorder (Sr-doping). This would mean that the local doping is so evenly distributed that there is no spatial variation that favor magnetic domain formation. A systematic study of this effect could be interesting.

\section{Conclusion}

In conclusion, by combining neutron diffraction, muon spin rotation, and magnetization measurements in strong magnetic fields, we have found that La$_{2-x}$Sr$_x$CuO$_{4+y}$ with $x=0.06$ shows no magnetic order at low temperature, and that the application of a magnetic field induces stripe ordered regions. The volume of these regions is proportional to the applied field value, while the ordered magnetic moment is field-independent. These findings are in contrast to the interpretation of earlier, similar data on field-induced magnetism in oxygen-stoichiometric \LSCO{} samples, where a field-induced enhancement of neutron diffraction intensity was interpreted to be caused by an increase in the ordered magnetic moment. Our results make it relevant to re-investigate with $\mu$SR whether the field-enhanced stripe signal in the other La-214 cuprates is due to an increase of the ordered moment, as previously concluded, or rather caused by an increased magnetic volume fraction. The answer to this is highly relevant for the understanding of the interplay between magnetism and SC in the cuprate superconductors.

When comparing with similar studies of other La$_{2-x}$Sr$_{x}$CuO$_{4+y}$ samples with different values of the Sr doping, $x$, we find that our value, $x=0.06$, presents the perfect balance of quenched disorder doping and the annealed ordered doping of mobile Oxygen in order to avoid static stripes from forming in zero magnetic field.
If the sketched equivalence of phenomenologies is correct, then what our sample provides is a clean laboratory for studying the competition between 2D superconducting, stripe-order (field induced) and 3D superconducting, stripe-free regions. 

\acknowledgements

This work is based on experiments performed at the Swiss Neutron Source SINQ and Muon Source S$\mu$S, Paul Scherrer Institute, Villigen, Switzerland. We also thank the Insitute Laue-Langevin (ILL), France for providing neutron beam time. 
We acknowledge DESY (Hamburg, Germany), a member of the Helmholtz Association HGF, for the provision of experimental facilities. Parts of this research were carried out at beamline BW5 at DORIS. 
We would like to thank Dr. Tatsuo Goko (PSI) for assistance during the $\mu$SR experiment and Prof. Dr. Kazimierz Conder (PSI) for performing TGA measurements.

The work has been supported by the Danish Foundation for Independent Research, by the Danish Agency for Science and Innovation through DANSCATT. SHD has been funded by NordForsk through NNSP, project 87865 and by the Carlsberg Foundation. HJ was funded by the EU Horizon 2020 program under the Marie Sklodowska-Curie grant agreement No 701647 and the Carlsberg Foundation. BOW was supported for sample production and data interpretation by the U.S. DOE-BES under Contract No. DE-FG02-00ER45801.

\bibliography{LSCObib.bib}

\clearpage
\newpage

\section*{APPENDIX A: Sample synthesis and characterization}

A large single crystal was grown at the Department of Energy Conversion and Storage, Technical University of Denmark, in a mirror furnace using the traveling solvent floating zone method. After cutting, we obtained a 20~mm long cylinder with a diameter of 10~mm and a mass of $m=7.9$~g. The nominal Sr content of the material (from the mixing of powders preceding the growth of the single crystal) was $x=0.06$. 

\begin{figure}[t]
\includegraphics[width=0.45\textwidth]{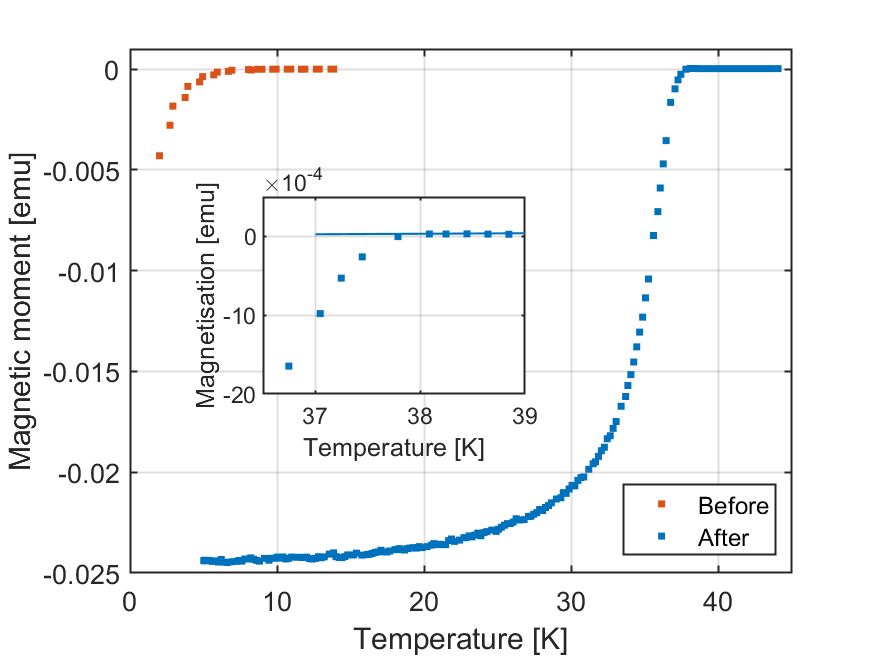} 
\caption{Magnetic moment measured before and after oxygenation in an applied field of 20 Oe. 
The insert is a zoom in of the data measured after oxygenation, from which we obtain the low-field onset critical temperature of $T_{\rm c} = 37.5(2)$~K.} \label{fig:PPMS}
\end{figure}

We measured the magnetic response of a small piece of the as-grown sample using an AC Magnetic Susceptometer set-up and found a diamagnetic signal below $\sim 6$~K, as shown with red dots in Fig.~\ref{fig:PPMS}. This value of the critical temperature is consistent with other Sr doped samples with $x \sim 0.06$ \cite{Fujita2002,Hawthorn2003}.  
The sample was oxygenated at the University of Connecticut through a wet-chemical technique for several months \cite{Mohottala2006}. The rod was cut into pieces to fit it in the cryo-magnets used in the neutron experiments. A small piece of oxygenated sample was measured with a VSM in an applied field of 20~Oe. It reveals a single transition for our La$_{1.94}$Sr$_{0.06}$CuO$_{4+y}$ sample, shown in Fig.~\ref{fig:PPMS} in blue dots.

At low temperatures the sample is orthorhombic, and we found $a\simeq b=5.33$~\AA, and $c = 13.15$~\AA{}. The most likely structure is low temperature orthorhombic, space group $Bmab$ \cite{Ray2017}, but multiple other space groups are possible depending on the exact nature of the oxygen octahedral tilts \cite{Bozin2015a}.

\subsection*{Oxygen mobility and cooling rates}
Excess oxygen has been shown to be mobile in La$_{2-x}$Sr$_{x}$CuO$_{4+y}$, and e.g.\ quench cooling of the sample from 200~K or 300~K yields very different low-temperature properties than after slow cooling \cite{Lorenz2002,Fratini2010}. 
For this reason, all our experiments reported were performed using a slow cooling rate (1~K/min or lower) from room temperature to 100~K in order to provide consistency between measurements, and to make sure the mobile excess oxygen ions find an optimal configuration in the crystal structure before freezing. 

\begin{figure*}
\centerline{
\includegraphics[width=0.95\textwidth]{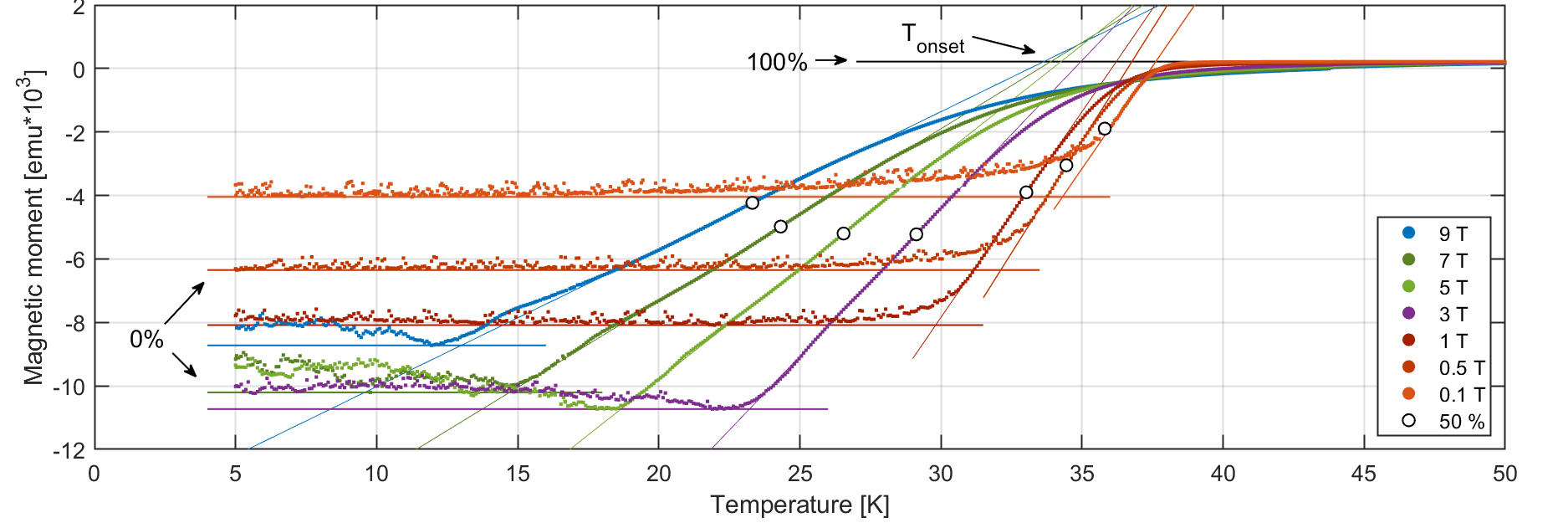}
}
\caption{Magnetic moment measured at different magnetic fields applied along the $c$-direction. The black line is a fit to the constant region above 50~K. The open circles indicate the 50\% magnetization point between the lowest point measured and the high-temperature constant value. The horizontal colored lines mark the lowest value obtained for each magnetization curve. Straight line fits are performed to the linear part of the slopes to obtain a value for $T_{\rm onset}$.
} \label{SC_rawMdata}
\end{figure*}

\section*{APPENDIX B: Magnetization measurements and the superconducting phase} 

In Fig.~\ref{SC_rawMdata}, we display the measured magnetization for different applied fields. 
The magnetic response of the sample at low temperatures is dominated by diamagnetism due to the superconducting phase for all applied fields in this study ($B \leq 9 $~T). 

The precise determination of a single $T_{\rm c}$ is made difficult by thermal fluctuations \cite{McGraw-Hill} that broaden the transition from the normal state to the superconducting state. We have therefore used two different methods to obtain a measure of $T_{\rm c}$ as a function of applied field. The first method is to find the halfway point between the full diamagnetic response and the normal state response \cite{VanBentum1987,Kwok1987}. The open circles in Fig.~\ref{SC_rawMdata} indicate this 50\% magnetization point for each field. In the second approach, we have performed a linear fit to the sloping part of the magnetization curves and found the intersection to the high temperature magnetization line as our onset temperature, $T_{\rm onset}$. In Fig.~\ref{fig:WHH}, both $T_{\rm onset}$ and the 50\% magnetization points are plotted as a function of applied magnetic field. 

\subsection*{Werthamer-Helfand-Hohenberg model}

The Werthamer-Helfand-Hohenberg theory for type-II superconductors can be used to estimate the upper critical field at zero temperature, $H_{c2}(0)$, from the slope of $H_{c2}$ as a function of temperature \cite{Werthamer1966}. For simplicity, we restrict our analysis to the dirty limit, where \cite{Werthamer1966,Nakamura2019} 
\begin{align}
    H_{c2}(0)=-0.69 T_c \frac{dH_{c2}}{dT}.
\end{align}
Typical values for $-dH_{c2}/{dT}$ in \LSCO{} are 0.5-1.5~T/K, e.g. Ref.~\onlinecite{Kwok1987}.

Using Ginzburg-Landau theory, the superconducting coherence length, $\xi$, can be estimated using \cite{Kittel2004}
\begin{align}
    H_{c2}(0)=\frac{\Phi_0}{2\pi \xi^2},
\end{align}
where $\Phi_0=2.06 \times 10^{-15}$ Wb.

\begin{figure}[h]
\includegraphics[width=0.45\textwidth]{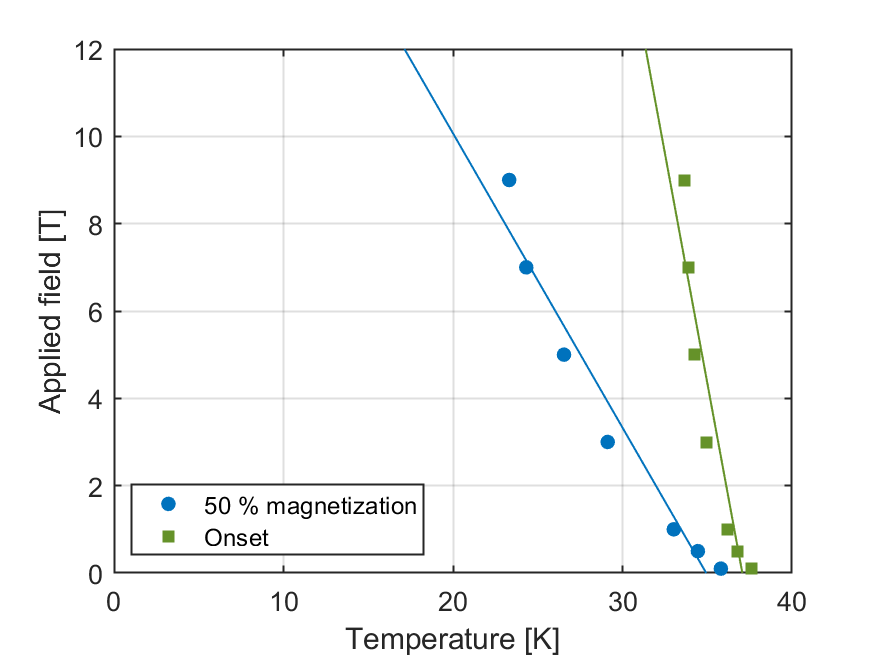}
\caption{$T_{\rm c}$ onset and 50\% magnetization point as a function of field, found from magnetization measurements in Fig.~\ref{SC_rawMdata}. It is the same data as displayed in Fig.~\ref{fig:temperature} (e) but here with the linear fits of the WHH model.
} \label{fig:WHH}
\end{figure}

Using the halfway point of the magnetization curves to estimate $H_{c2}$, we find $-{dH_{c2}}/{dT}=0.67(9)$~T/K and $T_{\rm c,50\%}=35.8$ K in our sample, giving $H_{c2}(0)=17(2)$ T and $\xi=4.5(6)$ nm.
Using the onset of diamagnetism as an estimate of $H_{c2}$, we find $-{dH_{c2}}/{dT}=2.1(3)$ T/K and $T_{\rm onset}=37.6$ K, and thus $H_{c2}(0)=54(8)$ T and $\xi=2.5(4)$ nm. This estimate of $H_{c2}(0)$ is likely an overestimate, as suggested by Ref.~\onlinecite{Nakamura2019}.
The lines in Fig.~\ref{fig:WHH} show the fits to the WHH model. 

We conclude that the superconducting coherence length is between 2.5-4.5 nm, assuming that the WHH model can describe our system. This estimate is, in itself, not self-consistent with the dirty limit approximation of the WHH model, where the superconducting coherence length should be larger than the mean free path of the electrons, which for \LSCO{} is of the order 10~nm \cite{Boyd2019}. For more precise results, a measurement of $H_{c2}(0)$ is required. Such measurements are outside the scope of this work. We note, however, that our estimates of $H_{c2}(0)$ are in agreement with measurements of $H_{c2}(0)$ in a broad range of LSCO samples \cite{Wang2008a}

\section*{APPENDIX C: Muon spin rotation} 

With the $\mu$SR technique, the time-dependent decay of an ensemble of muons is measured. The muons enter the sample, and in \LSCO{} they reside at one unique interstitial lattice site, which makes  $\mu$SR  a local probe that is sensitive to magnetic fields down to 10~$\mu$T inside the sample \cite{Blundell1999}.
The positive muons are initially completely spin polarized, and after implantation they  precess in the local magnetic field with a gyromagnetic ratio of $\gamma_{\mu} = 2\pi \times 135.5$~MHz/T. Muons are unstable particles, and decay into a positron and a neutrino-antineutrino pair with a lifetime of $\tau_{\mu} = 2.2 \, \mu$s. The positron is emitted preferentially along the direction of the muon spin, and the recorded decay positron count rate thus monitors the spin precession corresponding to the magnetic field strength at the muon site. The $\mu$SR spectra obtained in zero field conditions cover a time window of about $10^{-6}-10^{-9}$~s, which means that magnetic fluctuations faster than this are averaged to zero, while magnetic fluctuations slower than this are perceived as being static.
The average muon implantation depth is of the order 100~$\mu$m, and hence the magnetic and superconducting properties measured with muons are representative of the bulk sample.

We carried out two $\mu$SR experiments at the S$\mu$S facility at the Paul Scherrer Institute (CH), as described below.

\subsection*{Zero field $\mu$SR measurements}

A zero field experiment was performed on the GPS spectrometer. We used the forward (in front of the sample) and backward (behind the sample) detector with respect to the muon beam direction to derive the muon spin asymmetry, defined as
\begin{equation}
A\left(t\right) = \frac{N^{\rm{f}}\left(t\right) - \alpha N^{\rm{b}}\left(t\right)}{N^{\rm{f}}\left(t\right) + \alpha N^{\rm{b}}\left(t\right)},
\end{equation}
where $\alpha$ is a scaling factor to account for different efficiency and solid angle coverage of the two detectors.

%
\begin{figure}[b]
\includegraphics[width=0.4\textwidth]{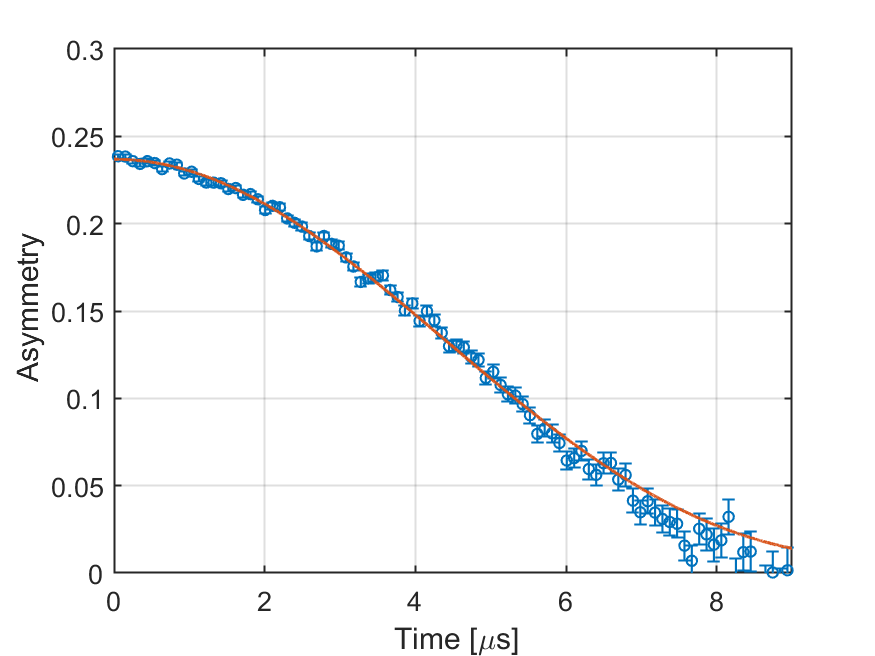}
\caption{$\mu$SR spectrum recorded under zero field conditions at 5~K. The red line is the result from a Gaussian Kubo-Toyabe-function fitted to the data.
} \label{fig:ZFdata}
\end{figure}

Fig.~\ref{fig:ZFdata} shows ZF-$\mu$SR data measured at 5~K, fitted by a Gaussian Kubo-Toyabe (KT) function:
\begin{equation}
A(t) = \tfrac{1}{3} + \tfrac{2}{3}(1-\sigma^2t^2)\exp(-\sigma^2 t^2 / 2).
\end{equation}
From this we find $\sigma= 0.168(1)~\mu s^{-1} \sim \gamma_{\mu} \Delta$, where $\Delta$ is the second moment of the magnetic field distribution at the muon site \cite{Hayano1979}.
A Gaussian KT function models the relaxation of muons stopping in a non-magnetic environment with static and randomly oriented nuclear moments. There is a slight deviation between data and model for long decay times ($>6$~$\mu$s), which indicates that the moments may be fluctuating slightly or that the system is not completely disordered, in agreement with the neutron diffraction data shown in the main paper.

\subsection*{High transverse field $\mu$SR measurements} 

In this work, we used the unique possibility of the HAL9500 instrument to apply a magnetic field up to 8~T to the sample. The instrument has 8 detectors in front of the sample position and 8 detectors behind. The muon precession phase, normalization, and background counts of each detector are fitted separately, while the key parameters of the model (asymmetry, relaxation rates, and oscillation frequency), are shared between all detectors. The positron count rate in each detector is therefore fitted to the following function
\begin{equation}
N^{i}\left(t\right) = N_{0}^{i} \exp{(-t/\tau_{\mu})} \left( 1 + A\left(t\right) \right) + N_{bg}^{i},
\end{equation}
where $N^{i}\left(t\right)$ is the number of positrons detected at time $t$ in detector $i$, and $N_{0}^{i}$ and $N_{bg}^{i}$ are the initial number of positrons and the background count in the detector, respectively. From the  asymmetry function, $A(t)$, physical parameters such as strength and distribution of the magnetic field within the sample can be extracted from fits to the models presented below. 

\begin{figure}[b]
\includegraphics[width=0.4\textwidth]{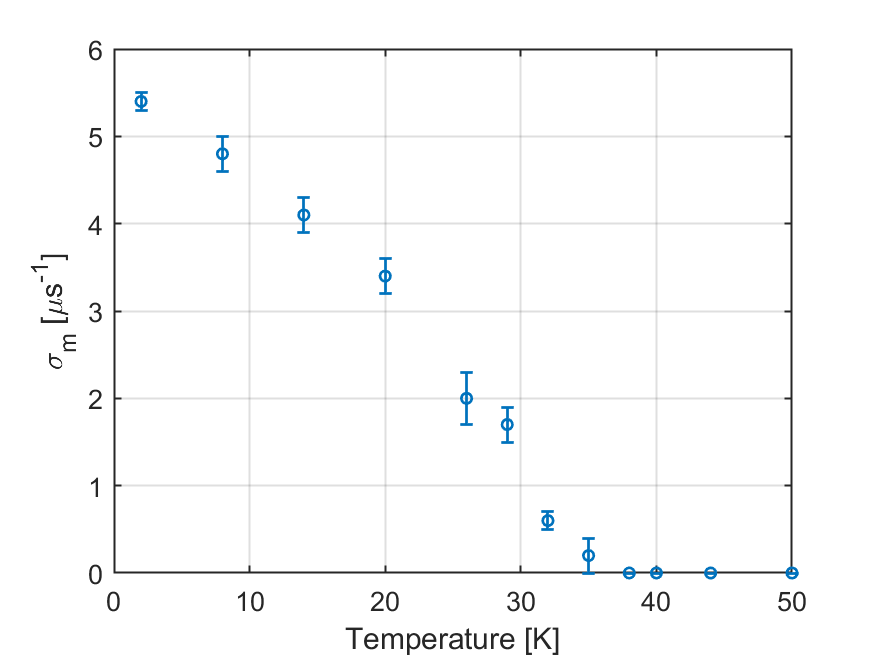} 
\caption{The relaxation rate of the magnetic regions of the sample as a function of temperature (see eq.~(\ref{twocomp})). The sample was  cooled in an applied field of 8~T.
} \label{fig:muSRresult_temp}
\end{figure}

The high transverse field $\mu$SR data has been fitted with a two-component model with the following function represented in a rotating reference frame;
\begin{align} \nonumber
A\left(t\right) &= V_{\rm{m}} \exp{\left( -\sigma_{\rm{m}}^2 t^2 /2 \right)} \cos \left( \omega_{\rm{m}} t + \phi^{i} \right) \\
&+ \left( 1 - V_{\rm{m}} \right) \exp{\left(-\sigma_{\rm{SC}}^2 t^2 /2 \right)} \cos \left( \omega_{\rm{SC}} t + \phi^{i} \right), \label{twocomp}
\end{align}
where $\omega_{\rm{m}}$ and $\omega_{\rm{SC}}$ are the Larmor frequencies of the muon spin and $\sigma_{\rm{m}}$ and $\sigma_{\rm{SC}}$ are the second moments of the magnetic field distribution at the muon site. We denote the magnetically ordered (superconducting) regions with the subscript $m$ (SC). The magnetic volume fraction of the sample is parameterized by $V_{\rm{m}}$. $\phi^{i}$ is the initial phase of the muon spin, determined by the detector geometry.

In general, a fast relaxation of the $\mu$SR spectrum due to magnetic correlations is analyzed using a model with an exponential envelope. However, for simplicity we have chosen a model where both signals are fitted with Gaussian components. This choice does not affect our results or their interpretation.

\begin{figure}[t]
\includegraphics[width=0.45\textwidth]{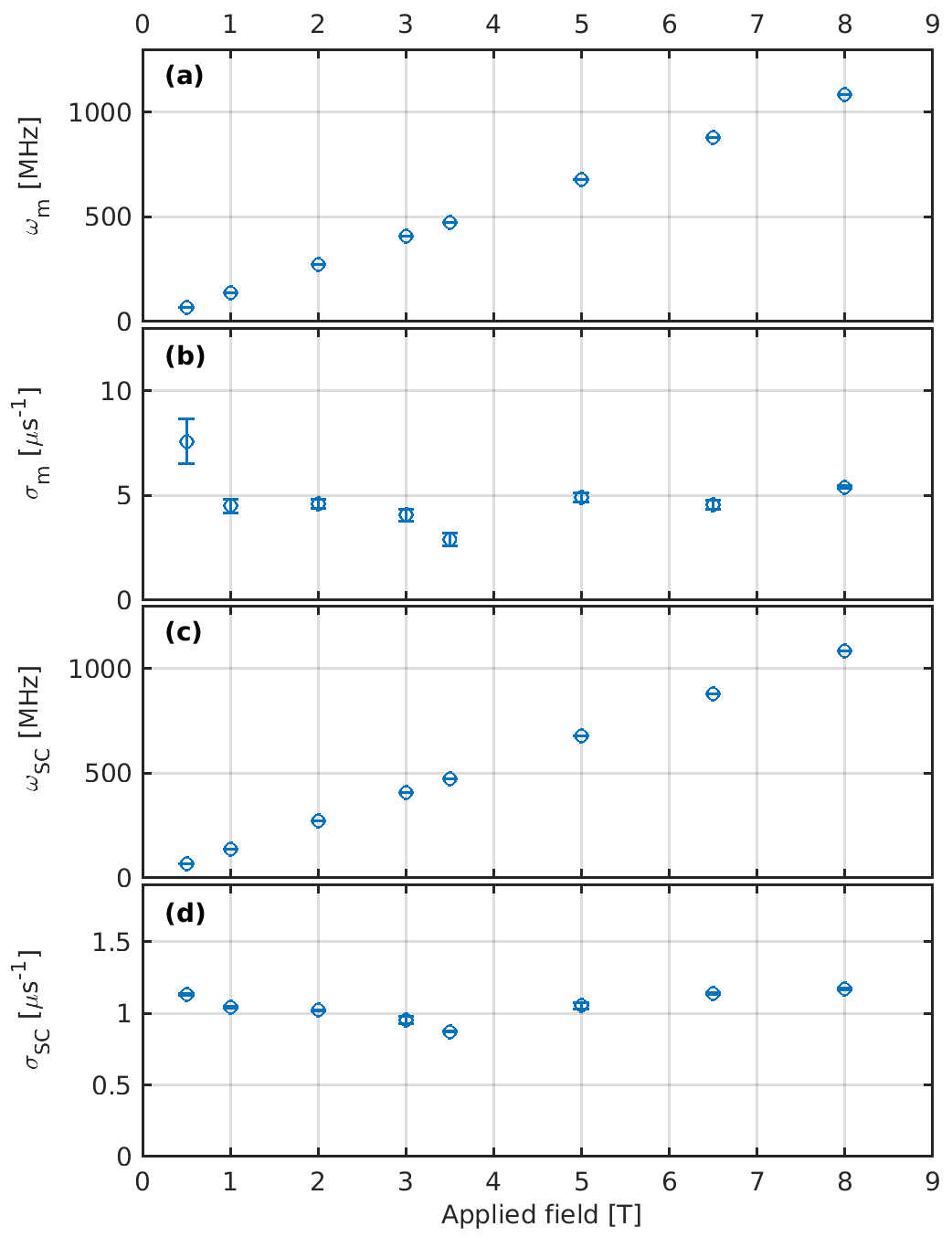} 
\caption{Parameters found from the two-component fits (Eq.~(\ref{twocomp})) as a function of applied field. A plot of $V_{\rm{m}}$ {\em vs.}\ field can be found in Fig.~\ref{fig:linear}. The sample was heated and field cooled to 2.5~K for each value of the field.
} \label{fig:muSRresult_field}
\end{figure}

A selection of raw data and fits, shown in rotating reference frames, can be found in Fig.~\ref{fig:rawdata} (b-d). We performed a temperature dependence study with an applied field of 8~T, and a field dependence study at 5~K.

In Fig.~\ref{fig:muSRresult_temp} the magnetic relaxation rate is shown as a function of temperature in an applied field of 8~T. The relaxation rate decreases as a function of temperature and reaches zero at $T_{\rm N}$.
The other three fitting parameters ($\sigma_{\rm{SC}}$, $\omega_{\rm{SC}}$, and $V_{\rm{m}}$) can be found in Fig.~\ref{fig:temperature} (b-d). $\omega_{\rm{m}}$ was held constant at 1084.63~MHz in the constant applied field. 

Fig.~\ref{fig:muSRresult_field} shows the fitting parameters as a function of applied field measured at 2.5~K. 
The rotation frequencies for the two regions of the sample, $\omega_{\rm{m}}$ and $\omega_{\rm{SC}}$, both display a linear field-dependence with a slope of 135.5 MHz/T as seen in Fig.~\ref{fig:muSRresult_field} (a, c). 
The magnetic relaxation rate at 2.5~K, $\sigma_{\rm{m}}$, shows a constant value close to 4.5 $\mu$s$^{-1}$ at high fields, but displays a slight upturn at low magnetic fields, Fig.~\ref{fig:muSRresult_field} (b). 
The superconducting relaxation rate, $\sigma_{\rm{SC}}$, varies little with field and has an average value close to 1~$\mu$s$^{-1}$. The small fluctuations around this value are caused by systematic uncertainties.

\section*{APPENDIX D: Neutron diffraction}

Neutron diffraction was performed at the cold-neutron triple axis spectrometers RITA-II at the Paul Scherrer Institute (PSI), Switzerland and ThALES at Institut Laue-Langevin (ILL), France. With a constant final energy of $5.0$~meV, the elastic energy resolution of the two experiments was close to 0.2~meV FWHM, while the $\mathbf{q}$-resolution was around 0.025~r.l.u. and 0.015~r.l.u.\ FWHM at ThaLES and RITA-II, respectively. 

RITA-II was configured in the monochromatic imaging mode \cite{bahlrita1,bahlrita2}, giving an effective collimation of 40' after the sample. No collimation was used before the sample.
No collimation was used on ThALES, but the analyzer geometry was unfocused (flat) and boron containing shielding was mounted to cover all but the central analyzer crystal, in order to enhance the ratio of signal-to-noise. In both experiments, a Be-filter was placed between sample and analyzer to filter out neutrons from second-order scattering. 

In the RITA-II experiment, the scans were performed along the (1+$h$, -$h$, 0)-direction in reciprocal space, which requires a small adjustment of scattering angle (A4) for each point of the scan. In contrast, in the ThALES experiment, we minimized the variations in background by doing pure sample-rotation (A3) scans. Here, the scan direction is along a (slightly) curved path in reciprocal space, but there is no change in the scattering angle (A4).

\subsection*{Structural modulation: Staging}

\begin{figure}[h]
\includegraphics[width=0.45\textwidth]{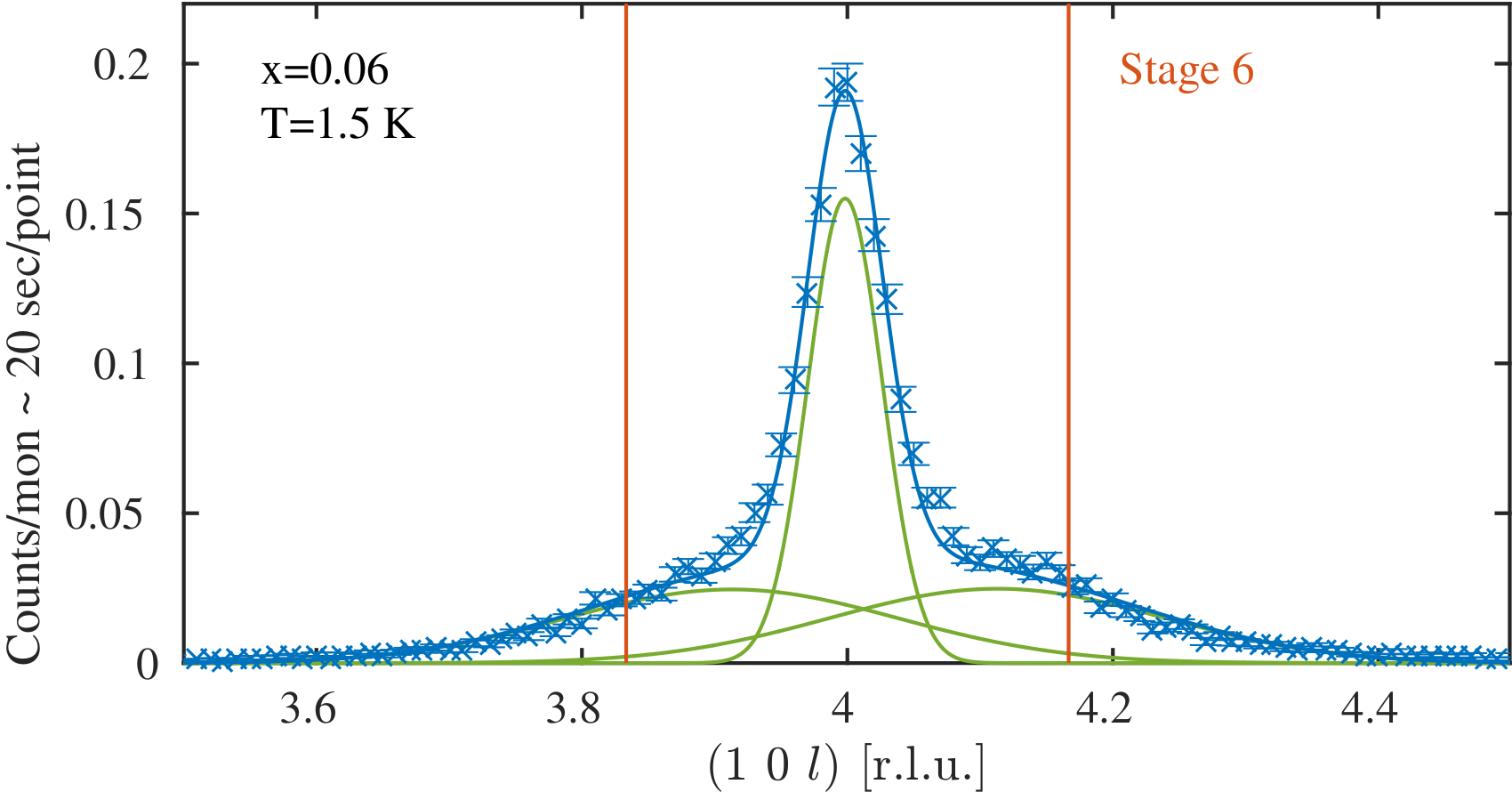}
\caption{Neutron diffraction scan along $l$, centered at the Bragg peak $\mathbf{q}=$(104). The signal from the ordering of the tilted octahedra appears on either side of the Bragg peak. The data is fitted by the sum of three Gaussians (blue line); one central peak and two broad satellites (green lines). The red lines indicate the position of the stage 6 peaks at $l=4\pm 1/6$. } \label{fig:Staging}
\end{figure}

The excess oxygen atoms cause anti-phase boundaries of the CuO$_6$ octahedra tilt pattern along the $c$-axis, usually denoted as “staging” due to the similarity of intercalated (staged) graphite \cite{Wells1996}. The periodicity of these tilt patterns can be observed as (pairs of) superstructure peaks when scanning through Bmab-allowed peaks along $l$, as shown in Fig.~\ref{fig:Staging}. Earlier, we have shown that the (main) staging peak position shifts closer to the Bmab position with increasing Sr content of the \LSCOO{} samples, indicating larger periodicity of the anti-phase tilts \cite{Ray2017}. Apart from the Bmab peak, we observe two broad peaks centered at $(0,1,4\pm 0.17)$ in our sample. The positions correspond to modulation of stage-6 or larger. The width of the peaks indicate a distribution of tilt-pattern periodicities with other stage values.

\end{document}